\documentclass[preprint]{aastex}

\usepackage{longtable}
\usepackage{natbib}

\shorttitle{Discordance of the unified scheme with observed properties}
\shortauthors{Singal}

\begin{document}

\title{Discordance of the unified scheme with observed properties of quasars and high-excitation galaxies in the 3CRR sample}

\author{Ashok K. Singal}
\affil{Astronomy and Astrophysics Division, Physical Research Laboratory,\\
Navrangpura, Ahmedabad - 380 009, India}
\email{asingal@prl.res.in} 
\begin{abstract}
We examine the consistency of the unified scheme of FR~II-type radio galaxies and quasars with their observed number and size distributions in 
the 3CRR sample. We separate the low-excitation galaxies from the high-excitation ones, as the former might not harbor a 
quasar within and thus may not be partaking in the unified scheme models. In the updated 3CRR sample, 
at low  redshifts ($z<0.5$), the relative number and luminosity distributions of high-excitation galaxies and quasars do roughly 
match the expectations from the orientation-based unified scheme model. However, a foreshortening in the observed 
sizes of quasars, a must in the orientation-based model, is not seen with respect to radio galaxies even when the low-excitation 
galaxies are excluded. This dashes the hope that the unified scheme might still work if one includes only the 
high-excitation galaxies. 
\end{abstract}
\section{Introduction}
In the prevalent orientation-based unified scheme (OUS),    
FR~II-type radio galaxies (RGs) and radio-loud quasars arise from the same
parent population of radio sources, and it is only the orientation with respect to the observer's line of sight that makes 
a source appear as a quasar or a radio galaxy (RG). It is thought that there is an optically thick torus of  
a half cone-opening angle $\xi_{c}$, that surrounds 
the nuclear continuum and the broad-line optical emission region, and that the axis of the torus coincides with the major radio-axis 
of the source. If the orientation of the source in the sky happens to be such that the observer's line of sight lies within the opening angle of the torus, 
then the observer is able to see the inner continuum and the broad-line region, and the source gets classified 
as a quasar. However if the observer's line of sight cuts through the obscuring toroidal region, then the observer cannot see the 
inner broad-line region, and the source gets classified as an RG. Thus it is the orientation of the source 
in the sky that decides whether it will appear as an RG or a quasar, the latter when the major radio-axis happens to be within a 
certain critical angle ($\xi_{c}$) around the observer's line of sight. RGs and quasars, in all other respects,  
are considered to be intrinsically the same. In this scenario, due to the foreshortening arising because of the sharper inclinations  
of their radio axes with respect to the observer's line of sight, 
the observed radio sizes of quasars should appear systematically smaller than those of RGs.

OUS has become quite popular because of its simplicity and the promise it holds to bring two apparently 
quite distinct class of objects, viz. quasars and RGs, under one roof. 
According to this scheme, it is the value of $\xi_{\rm c}$ that determines the relative numbers and  
sizes of quasars and RGs in a low-frequency radio-complete sample.  
Barthel (1989) pointed out that in the 3CRR sample, in the redshift range $0.5<z<1$, the
observed number as well as sizes of quasars are typically about half as large as those
of RGs. The corresponding value of $\xi_{\rm c} \sim 45^{\circ}$ has gained credence in the literature 
and is presently thought to be a `canonical' value for division between RGs and quasars.
However data in other redshift bins from the rest of the 3CRR sample 
do not seem to fit into this simple scenario as the relative number of RGs there far exceeds than that expected 
within the unified scheme (Singal 1993a). It was later pointed out (Laing et al. 1994) that at low redshifts,  
FR~II-type radio galaxies contain a significant number of low-excitation galaxies (LEGs) 
(see e.g. Hine \& Longair 1979), which are unlikely to appear
as quasars when seen end-on and should be excluded from the sample while testing the unified 
scheme models. Evidence for a population of powerful radio galaxies, concentrated at 
low redshifts, which lack the hidden quasar comes also from Infra-red observations (Ogle et al. 2006; Leipski et al. 2010). 
Hardcastle et al. (2009) from both X-ray and Mid-IR data, showed that almost all objects classed as LEGs
in optical spectroscopic studies lack a radiatively efficient active nucleus. Independently, from  
the observed opposite behavior of the luminosity--size correlations among RGs and quasars as well as the 
vast differences in their cosmological size evolutions, Singal (1988, 1993b, 1996) concluded that RGs and quasars 
could not be arising from the same parent population and therefore they do not fit in OUS type of models.

In samples selected at meter wavelengths, the flat-spectrum core-emission, if any, is highly suppressed with 
the relativistic beaming effects playing almost no part and the emission only from the steep 
spectrum extended parts of the source is observed. Therefore a comparison of the observed size of RGs and quasars 
is a very robust test, as at meter wavelengths both quasars and RGs are picked by the strength of only their extended radio emission, 
unaffected by any orientation effects. 
Since both RGs and quasars are supposed to arise from the same parent sample, 
there relative distributions should not depend upon on redshift or luminosity, and  
their observed size ratios should show the geometrical projection effects. 

The initial proposal for OUS was that all steep spectrum ($\alpha > 0.5$) 
FRII radio galaxies (RGs) and quasars belong to the same 
parent population, barring of course a small number of compact steep spectrum sources (CSSS, with size  
$l \stackrel{<}{_{\sim}}20$ kpc), which represent an altogether a different population of radio sources (Barthel 1989).  
According to this scheme every one of these RGs harbors a hidden quasar, which becomes apparent only when the major radio-axis 
of the source happens to be oriented within a certain critical angle ($\xi_{c}$) around the observer's line of sight. 
However, with the stipulation that LEGs do not contain a hidden quasar, the conventional wisdom is that 
HEGs alone partake  in unification with quasars and then it is expected that in this modified unification scheme, due to geometry, 
the observed numbers and sizes of quasars are expected to be smaller as compared to HEGs by factors based on the $\xi_{c}$ value. 
Such an investigation for comparison of HEGs and quasars numbers and sizes has not been done until now.

Here we attempt to verify whether the data in the updated 3CRR sample are indeed consistent with OUS, when the population  
of LEGs is excluded. This is in 
particular important because recently in an about five times deeper MRC sample (Kapahi et al. 1998a,b) an expected difference 
in radio sizes of quasar and RGs was not seen (Singal \& Singh 2013a) in any of the redshift bins, casting serious doubts on the 
unified scheme models. A similar inconsistency with OUS has been pointed out in a strong-source sample  
selected from the equatorial region of sky (Singal \& Singh 2013b), whose selection criteria are very similar to that of the 3CRR, 
and which presently is perhaps the best sample to match the 3CRR sample (Best et al. 1999). 
The pertinent question that needs an examination is -- after excluding LEGS, do we find the data in the 3CRR 
case consistent with OUS which was not seen in other independent sample? 
We critically scrutinize the 3CRR case in order to check its consistency with OUS.
\section{The 3CRR Sample}
Our chosen 3CRR sample (Laing et al. 1983) is radio complete in the sense that all radio sources stronger than its sensitivity 
limit $S_{178}=10.9$ Jy are included (and certainly with no spurious entries as each and every source in the sample 
has been studied in detail). 
The sample covers the sky north of declination, $\delta=10^\circ$, except for a zone of avoidance, a band of $\pm 10^\circ$ 
about the galactic plane. Also it has a 100\% optical identification content with detailed optical spectra to classify radio 
sources into radio galaxies and quasars. Much more detailed optical information has now become available for the 3CRR sample 
since papers by Barthel (1989) and Singal (1993a), 
with new data mostly on the recognition of low-excitation and high-excitation galaxies. Following Jackson \& Rawlings (1997), 
LEGs are defined as objects having (rest-frame) [O III] equivalent widths $<10$ \AA, [O II]/[O III] ratios $>1$, or both.
Also there are many more broad-line radio galaxies (BLRGs) or weak quasars (WQs), the latter with compact optical nuclei detected 
in infrared or X-rays in erstwhile classified narrow-line RGs (see Grimes et al. 2004). 
We have taken the updated optical and radio data from {\em https://www.astrosci.ca/users/willottc/3crr/3crr.html}. 
 
The 3CRR sample includes 8 sources (5 quasars and 3 RGs) with radio spectral index $\alpha \le 0.5$ (with $S\propto \nu^{-\alpha}$); 
we have excluded these flat-spectrum cases, 
since these mostly are core-dominant cases where the relativistic beaming might introduce serious selection effects.  
Only the strong FRII RGs (Fanaroff \& Riley 1974) with luminosity above a certain critical value are supposed to partake in unification 
with quasars. This critical luminosity value translates to $P_{408} > 5\times 10^{25}$ W Hz$^{-1}$ for more recent cosmological parameters 
$H_{0}=71\,$km~s$^{-1}$\,Mpc$^{-1}$, $\Omega_m=0.27$ and $\Omega_{\Lambda}=0.73$ 
(Spergel et al. 2003), and we have confined ourselves to sources only more luminous than that. 
It should be noted that the steep spectrum quasars and FRII RGs almost always have 
edge-brightened radio morphologies, which makes it possible to define size of the radio 
source between its extremities, independent of the sensitivity of the observing telescope.
There are 17 FRIs below this luminosity limit, which we have excluded;   
the quasars all except one fall above this luminosity limit. Also excluded are six FRI's 
(A1552, 3C288, 3C310, 3C314.1, 3C315, 3C346), which lie above this luminosity limit, but have the FRI radio morphology. 
Included among quasars there are 14 broad-line radio galaxies or weak quasars, all classified under WQs. 
Also there are seven compact steep spectrum sources 
(CSSS; linear size $< 20$ kpc), comprising 2 HEGs and 5 quasars, which seem to be a different class  
than the FRII class of sources whose unification is sought in OUS, and have therefore been excluded. 
Our final sample then contains 130 sources, with 85 RGs and 45 quasars. Among these 85 RGs there are 17 
LEGs, leaving 68 HEGs. 

Our sample differs from Barthel's (1989) sample (in the redshift range $0.5<z<1$) in that there are 3 WQs (3C22, 3C41, 3C325) 
which we have deleted from list of RGS and counted them among quasars. Also 3C345 has been excluded as it is a flat 
spectrum quasar ($\alpha \sim 0.27$). Thus we have 14 quasars instead of 12 as in Barthel's sample. 
Among RGs, 4C74.16 has been included as it fits 
the selection criteria, on the other hand 3C272 has been excluded as it does not meet the selection criteria of 3CRR sample (it flux-density 
falling below the sample criteria). Further as mentioned in Singal (1993a), the cases like that of 
3C226, 3C234 and 3C265, where the broad lines are seen in the polarized (reflected!) emission only (Grimes et al. 2004), 
need to be counted among the narrow-line radio galaxy population, insofar as the testing of the unified
scheme is concerned, since the obscuring torus surrounds their optical cores  and 
their radio axes are still assumed to be in the sky-plane.
In that sense even if it turns out later that some more galaxies in our sample show
broad lines in the polarized emission, they should not affect our results.
Thus there are 27 RGs (instead of 30 in Barthel's sample) out of which 26 are HEGs and one (3C427.1) LEG. 
Also note that LAS for 3C216 and 3C380 we have used are 3 and 7 arcsec respectively, the same as the values used by Barthel.

There are a total of 130 sources in our sample listed in Table 1, which presents IAU and 3CRR source name, 
flux-density $S_{408}$ at 408 MHz, the spectral index $\alpha$ ($S\propto \nu^{-\alpha}$), 
nature of optical object (HEG: High-excitation radio galaxy; LEG: Low-excitation radio galaxy; Q: quasar),
redshift $z$, largest angular size $\theta$ (in arcsec), linear size $l$ in kilo-pc and luminosity $P_{408}$ in W/Hz. 
The details of the formulation for calculating linear size and luminosity are described in Singal \& Singh (2013a).

\section{Results and discussion}
Figure 1(a) shows a histogram of the redshift ($z$) distributions of the low-excitation galaxies (LEGs) and high-excitation galaxies 
(HEGs), while Figure 1(b) shows that of quasars. The 
number of objects in each category are listed in either panel. The first thing that we note from the Figure 1(a) is that LEGs are indeed 
found only at low redshifts (with all 17 lying at $z<0.6$). 
Another thing that is apparent is that the quasar population at low redshifts (Figure 1(b)) comprises predominantly WQs. 
Figure 2(a) shows LEGS are also of relatively low luminosities as compared to HEGs, 
the latter peaking at a much higher luminosity value. Figure 2(b) shows the luminosity distribution of quasars, including WQs.
After excluding LEGs, the luminosity distribution of HEGs is quite similar to that of quasars (with WQs of course classified 
with quasars). 
 
Figure 3 shows normalized cumulative size distributions of RGs and quasars in various redshift bins. 
In the lowest redshift bin ($z<0.5$) there are 43 RGs and 13 quasars, the latter including 9 BLRGs (or WQs). 
RGs are divided into 16 LEGs and 27 HEGs, and are accordingly separately plotted in Figure 3(a). In Barthel's OUS model  
the number ratio of HEGs and quasars is expected to be close to two, and with the exclusion of LEGs from RGs 
(and of course inclusion of BLRGs or WQs along with quasars) we do get number ratios reasonably consistent with OUS, 
with quasar fraction $f_q = 13/(13+27) \sim 0.33$. 
At the same time a solution to the anomaly in the size distribution in the 3CRR sample in this redshift range ($z < 0.5$)  
is also sought by this exclusion of large number of LEGs, with relatively smaller radio sizes, to offset the 
large sizes of HEGs. In a picture consistent with Barthel's model, a quasar fraction ($\sim 0.33$) in this bin implies 
the sizes of quasar should be statistically smaller than those of the HEGs by a factor of about two. 
However, we do not find HEGs sizes to be larger than of quasar (Figure 3(a)). 
While it is true that LEGs do have systematically smaller size than HEGs, however their exclusion is still not sufficient 
to leave HEGs with median size value double than that of quasars as one would have expected in OUS, in 
consistency with their observed numbers.  

Table 2 shows a comparison of the observed ratio of the median size values with the expected ratio from
the unified-scheme model, calculated from the observed numbers of HEGs and 
quasars in different redshift bins. Table 2 presents the sub-sample used, 
number of quasars in that redshift bin, number of HEGs in that redshift bin, fraction of quasars in that redshift bin, 
median value (kpc) of size distribution for quasars, median value (kpc) of size distribution for HEGs,
critical angle $\xi_c$ (degrees) calculated from the observed number ratio of HEGs and quasars, 
expected size ratio $R_e$ of HEGs and quasars calculated from $\xi_c$ and the actually observed size ratio $R_o$. 

It should be noted that the median values are not very sensitive to errors in individual size measurements. 
The rms error in $l_{m}$ in each case is determined from the frequency distribution (histogram) of size distribution. 
The rms error is given by $\sqrt {n}/(2f_m)$ in units of the class interval of $l$ (Kendall 1945; Yule \& Kendall 1950), 
where $n$ is the total number of sources in that redshift bin and $f_m$ is the ordinate value of the smoothed frequency 
distribution at the median value in the respective histogram. 

From Figure 3(a) and Table 2 we see that quasars, if anything, appear to have sizes somewhat larger than of the 
HEGS in this lowest redshift bin ($z<0.5$); certainly they are not a factor of 1.7 smaller than the HEGs as would be 
expected from foreshortening due to geometric projection consistent with their number ratios. In fact from Table 2 we notice 
that the observed size ratio differs from that predicted by OUS from observed quasar fraction  at $3\sigma$ level. 
In no OUS scenario should quasar sizes be larger than of HEGs. Even in the limiting case when $\xi_c$ 
approaches $90^\circ$, the median value of quasar sizes distributed between $0^\circ$ and $90^\circ$ will be smaller than of 
the HEGs (all consequently in the sky plane) by a factor of $1/\sin(60^\circ)=1.15$. Thus the inconsistency of OUS 
with the 3CRR data in the $z<0.5$ bin, contrary to all expectations, does not get eliminated even when LEGs are 
excluded and a major difficulty for OUS in the low redshift region ($z<0.5$) still remains.
Thus with its basic tenet, that the observed quasar sizes should always be smaller than of the HEGs,   
having been violated, OUS seems to be thus discordant with the low redshift data. 

In the intermediate redshift range ($0.5 \le z <1$) the fraction of quasars is $14/(26+14) \sim 0.35$, 
and the quasar sizes also seem smaller than of the HEGs (Figure 3(b), Table 2). This of course should not come as a surprise  
since Barthel's (1989) proposal of OUS (with a cone opening angle $\xi_{c}\sim 45^{\circ}$), was based on the data in this very 
bin only. But strangely the data at high redshifts ($z>1$) too shows some anomaly, 
as even if the sizes of quasars are smaller with respect to those of HEGs there, in consistency with that 
the numbers of quasars is not smaller than of the HEGs, in fact their number looks to be even larger  (18 of them instead of 7
or 8 expected from 15 HEGs). It does not seem that OUS can be still supported from the 3CRR data. This is in addition 
to that all other samples (Singal \& Singh 2013a,b) have also shown incompatibility with OUS. 

Baldi et al. (2013) have examined the flux ratios of core fraction at 5 GHz with the integrated flux density at 178 MHz, 
and found it to be qualitatively consistent with OUS. It needs to be realized that core dominance may give only a 
qualitative discrimination for or against OUS, but it cannot provide quantitative tests as the 
core fraction have multivariable dependence, viz. orientation angle $\xi$ and the Lorentz factor $\gamma$. 
In addition to the intrinsic spread in the flux density 
of the core as well as of the extended emission, the uncertainties in spectral index between 178 MHz and 5 GHz could change 
the estimated flux of the extended component (or of the integrated emission) by more 
than an order of magnitude. Moreover for larger angles $\xi>1/\gamma$, the orientation is not very discriminating as 
flux of core does not vary steeply with angle. In fact for larger angles due to relativistic beaming the core flux rather 
decreases by a factor $\gamma$. 

We should clarify that the evidence against the unification of extended RGs and quasars here does not necessarily 
invalidate the relativistic beaming models (Orr \& Browne 1982) of the unification of core-dominated and 
lobe-dominated quasars. In the same way, any evidence seen 
in favor of the relativistic beaming models cannot be cited in favor of unification of extended RGs and quasars. 
The two unifications are independent even if these have been combined in the so-called grand unification  
scheme models of the active galactic nuclei (Urry \& Padovani 1995; Peterson 1997; 
Kembhavi \& Narlikar 1999). It  is quite likely that radio-loud quasars do not make a randomly oriented population;  
the question here is that do RGs and quasars fit together, as proposed by Barthel (1989), 
into one unified scheme model like OUS?

The results of Baldi et al. (2013) for the size tests are inconclusive as they 
find that the two size distributions do not differ at a $>90 \%$ level and they do not see the expected projection effects 
in case of quasars. We also find that though the number distributions do match with OUS after the LEGs are discarded especially 
for $z<0.5$ bin, which otherwise had shown a large excess of RGs with respect to quasars in Singal (1993a), however what we find still 
disconcerting is that the sizes of quasars mismatch the expected value due to projection at a $3\sigma$ level. 

Similar anomalies were seen in even other redshift bins in other samples that we had examined earlier (Singal \& Singh 2013a, 2013b).
Considering that in addition to the 3CRR, any other independent sample (MRC, BRL) that we examined earlier also showed absence of 
projection effects in quasar sizes, it seems that there is something amiss in the present 
version of the OUS. We might stress here that support for any scheme in terms of consistent data does not prove the theoretical 
model (since the data could be consistent with other very different models as well),  
however an evidence against does disprove the current model, 
at least the part that predicts smaller sizes for quasars. In that sense the data is discordant with the presently prevalent 
OUS models. 

One cannot save the situation by invoking a receding-torus-type scheme (Lawrence 1991; Hill et al. 1996) where the critical   
angle ($\xi_{c}$) may be evolving with redshift or luminosity, as no $\xi_{c}$ value will simultaneously satisfy both number and 
size ratios in the low ($z<0.5$) bin.
Thus the predictions of OUS are not corroborated by the data in low redshift bins $z<0.5$ even when LEGs have been excluded.

Except for that particular bin ($0.5 \le z <1$) of the 3CRR sample, which incidentally was instrumental in the proposition 
of the unified scheme with the ``canonical'' value $\xi_{\rm c}\sim 45^{\circ}$, data in other 
redshift bins do not seem to yield the expected size ratios. 
We have learnt that Boroson (2011; private communication) has constructed a new sample of high-luminosity extragalactic 
($0.1 < z < 0.5$) radio sources, using SDSS and three radio surveys -- WENSS, NVSS, and FIRST, taking care that the 
sample is not contaminated by anisotropic radio core emission. He found that the objects with broad lines 
(quasars) tend to have larger projected sizes than those without broad lines (RGs), and from that  he has 
argued that radio-loud quasars could not be unified with radio galaxies by  Barthel's (1989) orientation scheme, in agreement 
with our results. In a recent paper DiPompeo et al. (2013) re-examined size data in three different samples (including Barthel 1989, 
Boroson 2011 and Singal \& Singh 2013a) and come to conclusion that cast solely in terms of viewing angle effects, 
unification of these objections through orientation fails. Even allowing for a realistic intrinsic size distribution of RGs and quasars, 
DiPompeo et al. (2013) find 
that the results cannot easily be reconciled with the paradigm in which different projected sizes of radio sources result 
from projection only. They concluded that there is a significant probability that Barthel (1989) could find his claimed size ratio 
by chance even if there is no unification by orientation.
\section{Conclusion}
We showed that in the 3CRR sample, the observed sizes of quasar are not in any way systematically smaller than those 
of the high-excitation galaxies, even when the low-excitation radio galaxies are excluded, in the low redshift bin ($z<0.5$). 
Though an exclusion of the population of low-excitation radio galaxies, with apparently no hidden quasars inside,
could reduce the observed excess number of radio galaxies at low redshifts to make it in accordance with OUS, however it 
does not still change the size distribution sufficiently so as to make the sizes of the high-excitation galaxies larger 
than those of the quasars at these redshifts. 
The absence of this foreshortening of the sizes of the quasars as compared to that of the high-excitation galaxies 
is inconsistent with the unified scheme models. 
It looks like that Barthel's observation that sizes and numbers of quasars were smaller than of the RGs in $0.5 \le z<1$ bin 
was perhaps only a statistical fluctuation as a similar thing is not seen in the remainder of the 3CRR sample, implying 
that the 3CRR data is not in concordance with OUS.

\clearpage
\begin{figure}[h]
\scalebox{0.7}{\includegraphics{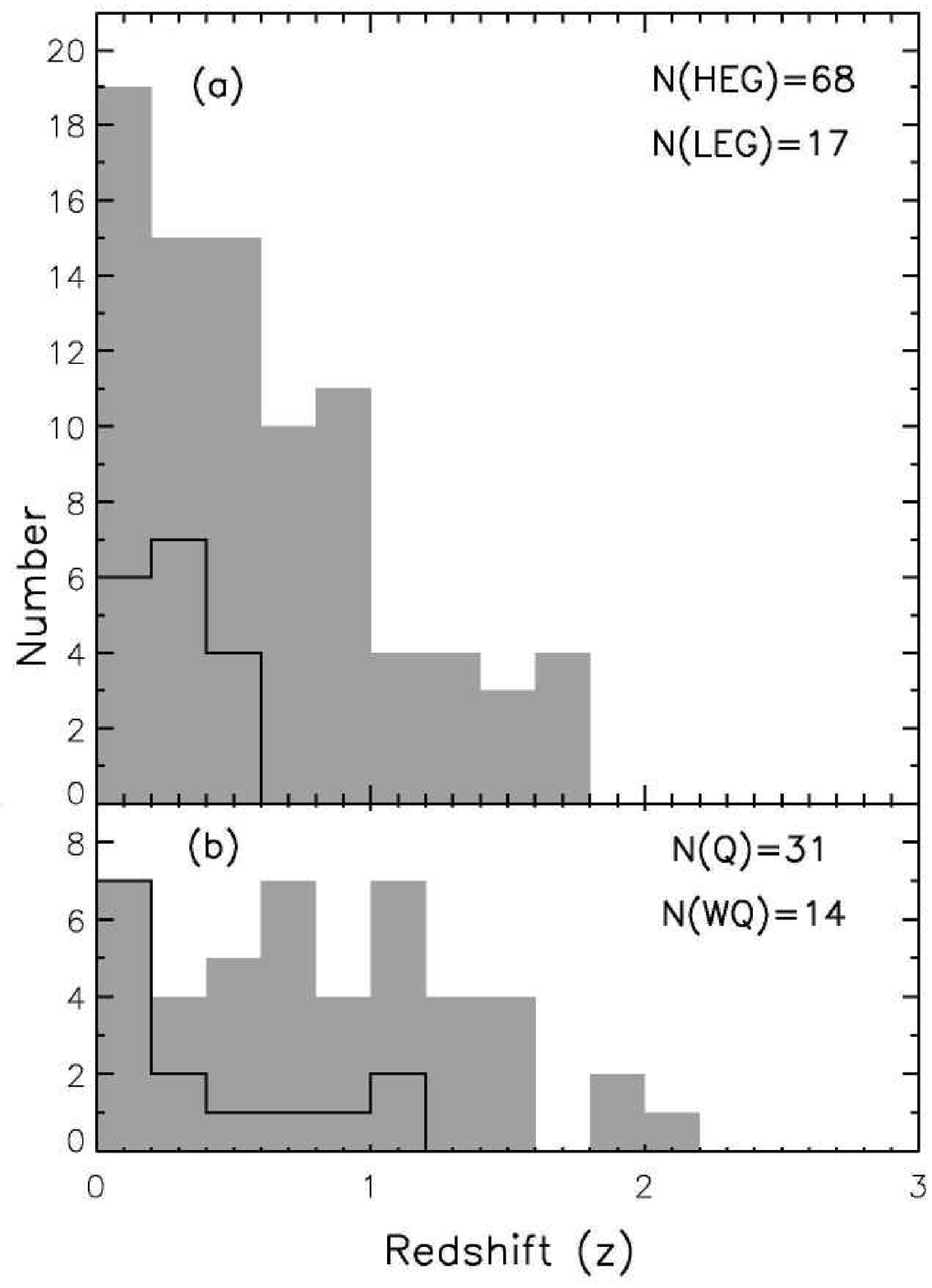}}
\caption{Histograms of the redshift ($z$) distributions of (a) FRII radio galaxies (shaded area) 
for the 3CRR sample with the low-excitation galaxies (LEGs) 
represented by the region under the overlaid darker lines, the remainder shaded region represents high-excitation galaxies 
(HEGs) (b) quasars (shaded area), the region under the overlaid dark line represents weak quasars (WQs) 
or BLRGs. Number of objects in various  categories are listed in each case.}
\end{figure}
\begin{figure}[h]
\scalebox{0.7}{\includegraphics{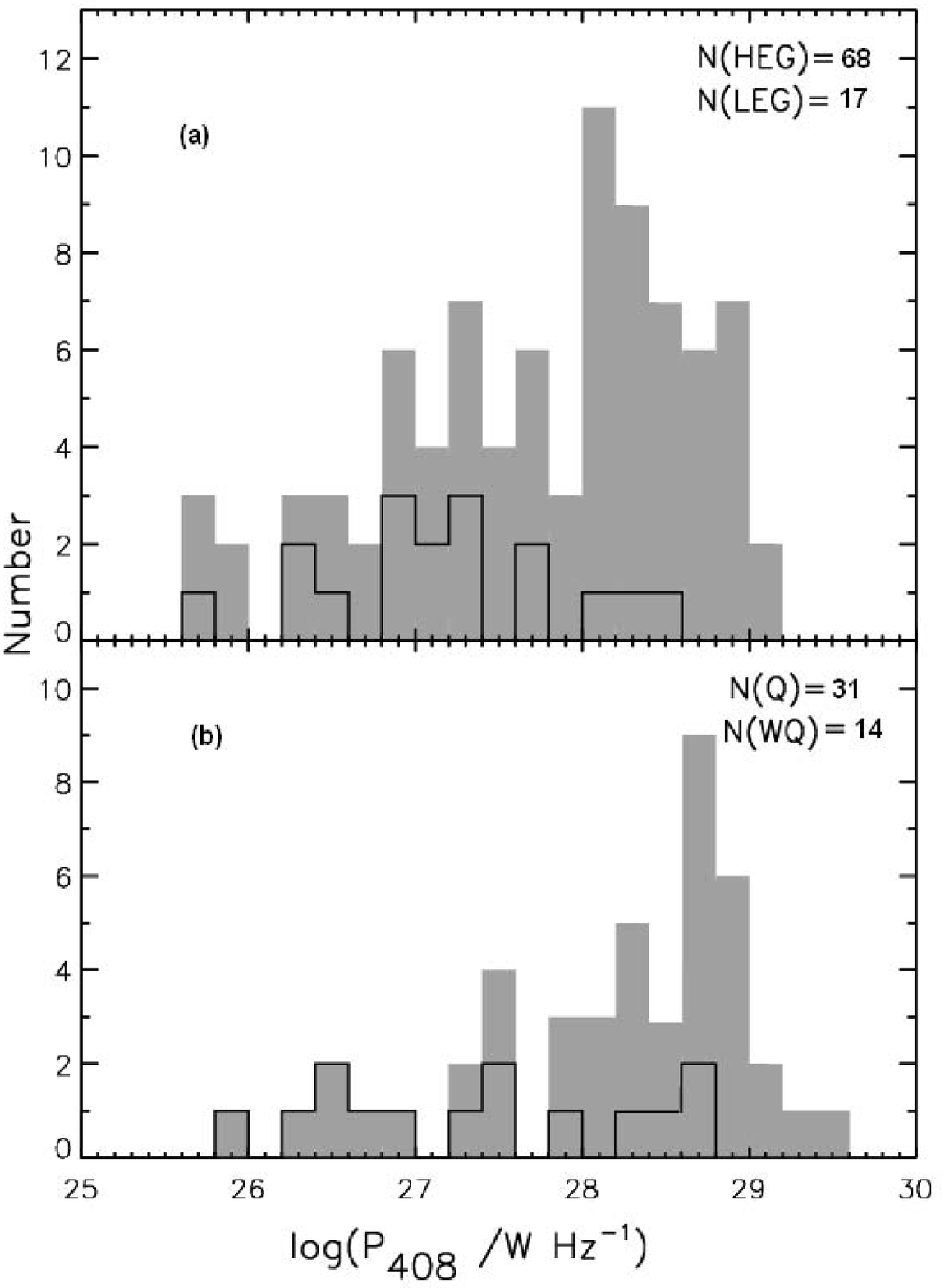}}
\caption{Histograms of the luminosity ($P_{408}$) distributions of (a) FRII radio galaxies (shaded area) 
for the 3CRR sample with the low-excitation galaxies (LEGs) 
represented by the region under the overlaid darker lines, the remainder shaded region represents high-excitation galaxies 
(HEGs) (b) quasars (shaded area), the region under the overlaid dark line represents weak quasars (WQs) 
or BLRGs. Number of objects in various  categories are listed in each case.}
\end{figure}
\begin{figure}[ht]
\scalebox{0.7}{\includegraphics{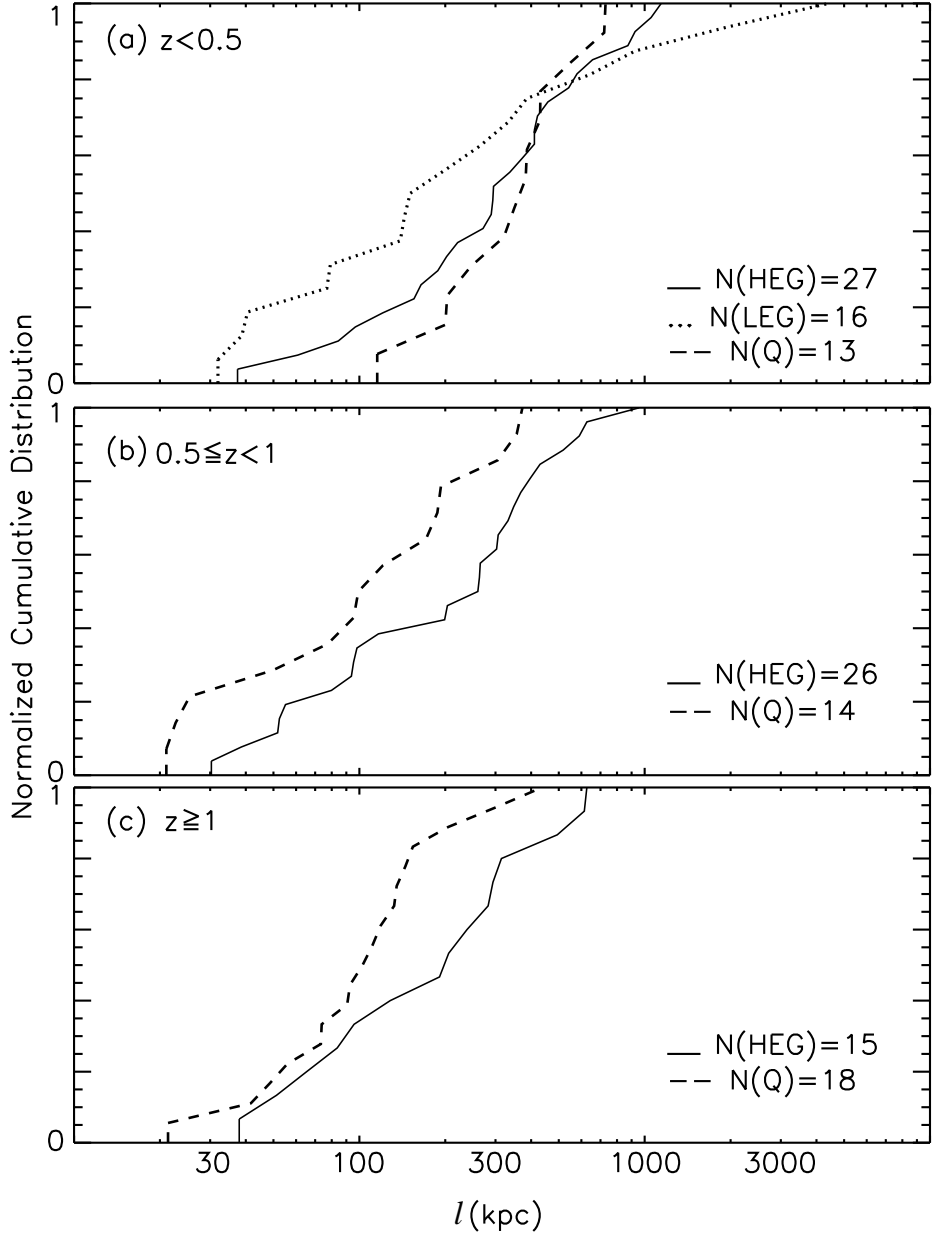}}
\caption{Normalized cumulative distributions of linear size ({\em l}) of HEGs (continuous curves), LEGs (dotted curves) 
 and quasars (dashed curves) in various redshift bins for the 3CRR sample. N(HEG), N(LEG) and N(Q) give the numbers of 
sources in each respective bin. Note that there is only one LEG at $z > 0.5$ in the 3CRR sample 
(3C427.1, $z=0.572$, $l= 183$ kpc), and which is not plotted in the figure.}
\end{figure}
\clearpage
\begin{longtable}{@{}clrcccrrc}
\caption{Radio and optical data for our sample.}\\
\hline\hline
IAU& \,3CRR & $S_{408}$ & $ \alpha$ & Opt & $z$ & $\theta\,\,$ &  $\l\,\,\, $ & $P_{408}$ \\
Name & Source  & Jy\,\, & & Obj & & \arcsec\,\,  & kpc & W Hz$^{-1}$\\
(1)&\,\,\,\,(2)&(3)&(4)&(5)&(6)&(7)&(8)&(9)\\
\hline\\
\endfirsthead
\tablename ~1\ -- \textit{continued} \\
\hline\hline
IAU& \,3CRR & $S_{408}$ & $ \alpha$ & Opt & $z$ & $\theta\,\,$ &  $\l\,\,\, $ & $P_{408}$ \\
Name & Source  & Jy\,\, & & Obj & & \arcsec\,\,  & kpc & W Hz$^{-1}$\\
(1)&\,\,\,\,(2)&(3)&(4)&(5)&(6)&(7)&(8)&(9)\\
\hline\\
\endhead
0007+124	&	4C12.03	&	5.3	&	0.9	&	LEG	&	0.16	&	240	&	642	&	3E+26	\\	
0013+790	&	3C6.1	&	8.5	&	0.7	&	HEG	&	0.84	&	26	&	199	&	2E+28	\\	
0017+154	&	3C9	&	7.7	&	1.1	&	Q	&	2.01	&	14	&	119	&	3E+29	\\	
0031+391	&	3C13	&	6.0	&	0.9	&	HEG	&	1.35	&	28	&	238	&	6E+28	\\	
0033+183	&	3C14	&	5.8	&	0.8	&	Q	&	1.47	&	24	&	205	&	7E+28	\\	
0035+130	&	3C16	&	5.6	&	0.9	&	HEG	&	0.41	&	78	&	421	&	3E+27	\\	
0038+328	&	3C19	&	7.8	&	0.6	&	LEG	&	0.48	&	7	&	41	&	6E+27	\\	
0040+517	&	3C20	&	27.0	&	0.7	&	HEG	&	0.17	&	53	&	155	&	2E+27	\\	
0048+509	&	3C22	&	6.9	&	0.8	&	WQ	&	0.94	&	24	&	193	&	3E+28	\\	
0053+261	&	3C28	&	7.4	&	1.1	&	LEG	&	0.20	&	43	&	139	&	8E+26	\\	
0106+130	&	3C33	&	31.6	&	0.8	&	HEG	&	0.06	&	257	&	290	&	3E+26	\\	
0106+729	&	3C33.1	&	8.5	&	0.6	&	WQ	&	0.18	&	239	&	721	&	7E+26	\\	
0107+315	&	3C34	&	5.4	&	1.1	&	HEG	&	0.69	&	47	&	332	&	1E+28	\\	
0109+492	&	3C35	&	6.0	&	0.8	&	LEG	&	0.07	&	730	&	926	&	6E+25	\\	
0123+329	&	3C41	&	7.6	&	0.5	&	WQ	&	0.79	&	25	&	188	&	2E+28	\\	
0125+287	&	3C42	&	7.1	&	0.7	&	HEG	&	0.40	&	31	&	165	&	4E+27	\\	
0127+233	&	3C43	&	6.8	&	0.8	&	Q	&	1.47	&	3	&	21	&	7E+28	\\	
0132+376	&	3C46	&	4.4	&	1.1	&	HEG	&	0.44	&	164	&	925	&	3E+27	\\	
0133+207	&	3C47	&	12.8	&	1.0	&	Q	&	0.43	&	78	&	431	&	8E+27	\\	
0154+286	&	3C55	&	9.9	&	1.0	&	HEG	&	0.74	&	71	&	518	&	2E+28	\\	
0210+860	&	3C61.1	&	18.0	&	0.8	&	HEG	&	0.19	&	186	&	579	&	2E+27	\\	
0220+397	&	3C65	&	8.9	&	0.8	&	WQ	&	1.18	&	17	&	145	&	6E+28	\\	
0229+341	&	3C68.1	&	7.2	&	0.8	&	Q	&	1.24	&	52	&	437	&	5E+28	\\	
0231+313	&	3C68.2	&	4.6	&	1.1	&	HEG	&	1.58	&	22	&	191	&	8E+28	\\	
0307+169	&	3C79	&	15.5	&	0.9	&	WQ	&	0.26	&	89	&	349	&	3E+27	\\	
0356+102	&	3C98	&	26.9	&	0.8	&	HEG	&	0.03	&	308	&	188	&	6E+25	\\	
0410+110	&	3C109	&	11.6	&	0.9	&	WQ	&	0.31	&	96	&	430	&	3E+27	\\	
0411+141	&	4C14.11	&	6.0	&	0.8	&	LEG	&	0.21	&	115	&	386	&	7E+26	\\	
0433+295	&	3C123	&	115.3	&	0.7	&	LEG	&	0.22	&	41	&	144	&	1E+28	\\	
0453+227	&	3C132	&	8.5	&	0.7	&	LEG	&	0.21	&	22	&	77	&	1E+27	\\	
0605+480	&	3C153	&	9.6	&	0.7	&	LEG	&	0.28	&	9	&	39	&	2E+27	\\	
0651+542	&	3C171	&	10.3	&	0.9	&	HEG	&	0.24	&	10	&	37	&	2E+27	\\	
0659+253	&	3C172	&	8.1	&	0.9	&	HEG	&	0.52	&	101	&	627	&	8E+27	\\	
0702+749	&	3C173.1	&	8.1	&	0.9	&	LEG	&	0.29	&	61	&	265	&	2E+27	\\	
0710+118	&	3C175	&	8.5	&	1.0	&	Q	&	0.77	&	48	&	356	&	2E+28	\\	
0711+146	&	3C175.1	&	5.8	&	0.9	&	HEG	&	0.92	&	7	&	55	&	2E+28	\\	
0725+147	&	3C181	&	6.9	&	1.0	&	Q	&	1.38	&	6	&	48	&	8E+28	\\	
0733+705	&	3C184	&	7.0	&	0.9	&	HEG	&	0.99	&	5	&	39	&	3E+28	\\	
0734+805	&	3C184.1	&	8.1	&	0.7	&	WQ	&	0.12	&	182	&	387	&	3E+26	\\	
0758+143	&	3C190	&	7.6	&	0.9	&	Q	&	1.20	&	7	&	56	&	6E+28	\\	
0802+103	&	3C191	&	6.3	&	1.0	&	Q	&	1.96	&	5	&	42	&	2E+29	\\	
0802+243	&	3C192	&	11.9	&	0.8	&	HEG	&	0.06	&	196	&	221	&	1E+26	\\	
0809+483	&	3C196	&	38.6	&	0.8	&	Q	&	0.87	&	10	&	77	&	1E+29	\\	
0824+294	&	3C200	&	6.1	&	0.8	&	LEG	&	0.46	&	26	&	151	&	4E+27	\\	
0832+143	&	4C14.27	&	4.3	&	1.2	&	LEG	&	0.39	&	38	&	201	&	2E+27	\\	
0833+654	&	3C204	&	4.7	&	1.1	&	Q	&	1.11	&	37	&	302	&	3E+28	\\	
0835+580	&	3C205	&	6.6	&	0.9	&	Q	&	1.53	&	18	&	154	&	9E+28	\\	
0838+133	&	3C207	&	7.0	&	0.9	&	Q	&	0.68	&	14	&	99	&	1E+28	\\	
0850+140	&	3C208	&	8.3	&	1.0	&	Q	&	1.11	&	11	&	91	&	5E+28	\\	
0855+143	&	3C212	&	7.7	&	0.9	&	Q	&	1.05	&	9	&	73	&	4E+28	\\	
0903+169	&	3C215	&	5.2	&	1.1	&	Q	&	0.41	&	59	&	321	&	3E+27	\\	
0905+380	&	3C216	&	11.6	&	0.8	&	Q	&	0.67	&	3	&	21	&	2E+28	\\	
0906+430	&	3C217	&	6.1	&	0.8	&	HEG	&	0.90	&	12	&	94	&	2E+28	\\	
0917+458	&	3C219	&	22.9	&	0.8	&	WQ	&	0.17	&	189	&	553	&	2E+27	\\	
0926+793	&	3C220.1	&	8.0	&	0.9	&	HEG	&	0.62	&	30	&	204	&	1E+28	\\	
0931+836	&	3C220.3	&	9.2	&	0.8	&	HEG	&	0.69	&	7	&	52	&	2E+28	\\	
0936+361	&	3C223	&	8.7	&	0.7	&	WQ	&	0.14	&	306	&	729	&	4E+26	\\	
0939+139	&	3C225B	&	10.6	&	0.9	&	HEG	&	0.58	&	5	&	30	&	1E+28	\\	
0941+100	&	3C226	&	7.9	&	0.9	&	HEG	&	0.82	&	35	&	265	&	2E+28	\\	
0945+734	&	4C73.08	&	7.7	&	0.9	&	HEG	&	0.06	&	947	&	1053	&	6E+25	\\	
0947+145	&	3C228	&	10.4	&	1.0	&	HEG	&	0.55	&	47	&	303	&	1E+28	\\	
0958+290	&	3C234	&	16.8	&	0.9	&	HEG	&	0.18	&	110	&	337	&	2E+27	\\	
1003+351	&	3C236	&	10.3	&	0.5	&	LEG	&	0.10	&	2440	&	4408	&	2E+26	\\	
1008+467	&	3C239	&	5.9	&	1.1	&	HEG	&	1.78	&	11	&	96	&	1E+29	\\	
1009+748	&	4C74.16	&	6.2	&	0.9	&	HEG	&	0.57	&	40	&	260	&	8E+27	\\	
1030+585	&	3C244.1	&	11.2	&	0.8	&	HEG	&	0.43	&	53	&	295	&	7E+27	\\	
1040+123	&	3C245	&	8.2	&	0.8	&	Q	&	1.03	&	9	&	74	&	4E+28	\\	
1056+432	&	3C247	&	7.0	&	0.6	&	HEG	&	0.75	&	13	&	95	&	1E+28	\\	
1100+772	&	3C249.1	&	6.0	&	0.8	&	Q	&	0.31	&	44	&	200	&	2E+27	\\	
1108+359	&	3C252	&	5.1	&	1.0	&	HEG	&	1.10	&	60	&	494	&	4E+28	\\	
1111+408	&	3C254	&	9.8	&	1.0	&	Q	&	0.73	&	13	&	95	&	2E+28	\\	
1137+660	&	3C263	&	8.4	&	0.8	&	Q	&	0.65	&	44	&	305	&	1E+28	\\	
1140+223	&	3C263.1	&	9.6	&	0.9	&	HEG	&	0.82	&	7	&	52	&	3E+28	\\	
1142+318	&	3C265	&	9.6	&	1.0	&	HEG	&	0.81	&	78	&	590	&	3E+28	\\	
1143+500	&	3C266	&	5.2	&	1.0	&	HEG	&	1.28	&	5	&	38	&	5E+28	\\	
1147+130	&	3C267	&	7.4	&	0.9	&	HEG	&	1.14	&	38	&	315	&	5E+28	\\	
1157+732	&	3C268.1	&	14.3	&	0.6	&	HEG	&	0.97	&	46	&	368	&	5E+28	\\	
1206+439	&	3C268.4	&	5.8	&	0.8	&	Q	&	1.40	&	11	&	93	&	6E+28	\\	
1218+339	&	3C270.1	&	8.0	&	0.8	&	Q	&	1.52	&	12	&	103	&	9E+28	\\	
1232+216	&	3C274.1	&	8.7	&	0.9	&	HEG	&	0.42	&	158	&	873	&	5E+27	\\	
1241+166	&	3C275.1	&	9.0	&	1.0	&	Q	&	0.56	&	19	&	121	&	1E+28	\\	
1251+159	&	3C277.2	&	5.6	&	1.0	&	HEG	&	0.77	&	58	&	430	&	2E+28	\\	
1254+476	&	3C280	&	13.2	&	0.8	&	HEG	&	1.00	&	15	&	117	&	6E+28	\\	
1308+277	&	3C284	&	5.6	&	1.0	&	HEG	&	0.24	&	175	&	657	&	9E+26	\\	
1319+428	&	3C285	&	5.6	&	1.0	&	HEG	&	0.08	&	184	&	271	&	8E+25	\\	
1343+500	&	3C289	&	6.7	&	0.8	&	HEG	&	0.97	&	10	&	80	&	3E+28	\\	
1349+647	&	3C292	&	5.7	&	0.8	&	HEG	&	0.71	&	133	&	958	&	1E+28	\\	
1404+344	&	3C294	&	4.6	&	1.1	&	HEG	&	1.79	&	15	&	128	&	1E+29	\\	
1409+524	&	3C295	&	54.0	&	0.6	&	LEG	&	0.46	&	5	&	32	&	4E+28	\\	
1419+419	&	3C299	&	7.5	&	0.7	&	HEG	&	0.37	&	12	&	61	&	3E+27	\\	
1420+198	&	3C300	&	10.2	&	0.8	&	HEG	&	0.27	&	100	&	411	&	2E+27	\\	
1441+522	&	3C303	&	6.5	&	0.8	&	WQ	&	0.14	&	47	&	116	&	3E+26	\\	
1458+718	&	3C309.1	&	15.9	&	0.5	&	Q	&	0.90	&	3	&	23	&	5E+28	\\	
1522+546	&	3C319	&	7.9	&	0.9	&	LEG	&	0.19	&	105	&	333	&	8E+26	\\	
1529+242	&	3C321	&	8.9	&	0.6	&	HEG	&	0.10	&	308	&	542	&	2E+26	\\	
1533+557	&	3C322	&	5.6	&	0.8	&	HEG	&	1.68	&	33	&	283	&	9E+28	\\	
1547+215	&	3C324	&	8.2	&	0.9	&	HEG	&	1.21	&	10	&	84	&	6E+28	\\	
1549+202	&	3C325	&	8.2	&	0.9	&	WQ	&	1.14	&	16	&	132	&	5E+28	\\	
1549+628	&	3C326	&	12.4	&	0.7	&	LEG	&	0.09	&	1190	&	1937	&	2E+26	\\	
1609+660	&	3C330	&	16.8	&	0.7	&	HEG	&	0.55	&	62	&	397	&	2E+28	\\	
1618+177	&	3C334	&	5.8	&	0.9	&	Q	&	0.56	&	58	&	373	&	7E+27	\\	
1622+238	&	3C336	&	6.8	&	0.7	&	Q	&	0.93	&	22	&	171	&	2E+28	\\	
1626+278	&	3C337	&	6.4	&	0.9	&	HEG	&	0.64	&	45	&	307	&	1E+28	\\	
1627+234	&	3C340	&	6.0	&	0.7	&	HEG	&	0.78	&	47	&	348	&	1E+28	\\	
1627+444	&	3C341	&	7.0	&	0.6	&	HEG	&	0.45	&	80	&	458	&	4E+27	\\	
1658+471	&	3C349	&	7.8	&	0.7	&	HEG	&	0.21	&	88	&	294	&	9E+26	\\	
1704+608	&	3C351	&	8.1	&	0.7	&	Q	&	0.37	&	75	&	383	&	3E+27	\\	
1709+460	&	3C352	&	5.9	&	0.9	&	HEG	&	0.81	&	13	&	98	&	2E+28	\\	
1723+510	&	3C356	&	5.3	&	1.0	&	HEG	&	1.08	&	75	&	614	&	3E+28	\\	
1732+160	&	4C16.49	&	5.0	&	1.0	&	Q	&	1.30	&	16	&	135	&	5E+28	\\	
1759+137	&	4C13.66	&	6.3	&	0.8	&	HEG	&	1.45	&	6	&	51	&	7E+28	\\	
1802+110	&	3C368	&	5.4	&	1.2	&	HEG	&	1.13	&	8	&	65	&	5E+28	\\	
1828+487	&	3C380	&	35.9	&	0.7	&	Q	&	0.69	&	7	&	50	&	6E+28	\\	
1832+474	&	3C381	&	9.2	&	0.8	&	HEG	&	0.16	&	74	&	202	&	6E+26	\\	
1833+326	&	3C382	&	13.3	&	0.6	&	WQ	&	0.06	&	186	&	203	&	1E+26	\\	
1842+455	&	3C388	&	15.0	&	0.7	&	HEG	&	0.09	&	51	&	84	&	3E+26	\\	
1845+797	&	3C390.3	&	27.8	&	0.8	&	WQ	&	0.06	&	229	&	246	&	2E+26	\\	
1939+605	&	3C401	&	12.6	&	0.7	&	LEG	&	0.20	&	24	&	79	&	1E+27	\\	
2104+763	&	3C427.1	&	13.0	&	1.0	&	LEG	&	0.57	&	28	&	183	&	2E+28	\\	
2120+168	&	3C432	&	5.3	&	1.0	&	Q	&	1.81	&	13	&	111	&	1E+29	\\	
2121+248	&	3C433	&	32.9	&	0.8	&	HEG	&	0.10	&	66	&	121	&	8E+26	\\	
2141+279	&	3C436	&	9.5	&	0.9	&	HEG	&	0.21	&	108	&	372	&	1E+27	\\	
2145+151	&	3C437	&	8.3	&	0.8	&	HEG	&	1.48	&	34	&	294	&	1E+29	\\	
2153+377	&	3C438	&	23.5	&	0.9	&	HEG	&	0.29	&	22	&	97	&	6E+27	\\	
2203+292	&	3C441	&	6.9	&	0.8	&	HEG	&	0.71	&	37	&	264	&	1E+28	\\	
2243+394	&	3C452	&	31.0	&	0.8	&	HEG	&	0.08	&	272	&	411	&	5E+26	\\	
2252+129	&	3C455	&	7.7	&	0.7	&	WQ	&	0.54	&	4	&	25	&	8E+27	\\	
2309+184	&	3C457	&	6.2	&	1.0	&	HEG	&	0.43	&	205	&	1142	&	4E+27	\\	
2352+796	&	3C469.1	&	5.5	&	1.0	&	HEG	&	1.34	&	74	&	627	&	6E+28	\\	
2356+438	&	3C470	&	5.8	&	0.8	&	HEG	&	1.65	&	24	&	206	&	9E+28	\\	
\end{longtable}
\begin{table}[ht]
\caption{The observed ratios of the median size values of HEGs and quasars
compared with the expected ratio calculated from their relative numbers in various redshift bins.}
\begin{tabular}{@{\hspace{0mm}}c@{\hspace{3mm}}c@{\hspace{4mm}}c@{\hspace{4mm}}c@{\hspace{4mm}}c
@{\hspace{4mm}}c@{\hspace{4mm}}c@{\hspace{4mm}}c@{\hspace{4mm}}c@{\hspace{4mm}}c}\\
\hline\hline
Sub-sample & N(Q) & N(HEG) & $f_q$ & $l_{m}$(Q) & $l_{m}$(HEG)& $\xi_{c}$ & $R_{e}$ & $R_{o}$\\
(1)&\,(2)&(3)&(4)&(5)&(6)&(7)&(8)&(9)\\
\hline
& & & & & & & & \\
$z<0.5$ & 13 & 27 & 0.33 & $382\stackrel{+62}{_{-54}}$ & $295\stackrel{+65}{_{-54}}$& 47.5 & 1.7 & $0.8\stackrel{+0.3}{_{-0.2}}$\\\\
$0.5\leq z<1$ & 14 & 26 & 0.35 & $110\stackrel{+36}{_{-27}}$  & $262\stackrel{+47}{_{-40}}$& 49.5 & 1.7 & $2.4\stackrel{+0.9}{_{-0.7}}$\\\\
$z>1$ & 18 & 15 & 0.55 & $107\stackrel{+26}{_{-21}}$& $206\stackrel{+60}{_{-46}}$ & 63.0 & 1.4 & $1.9\stackrel{+0.8}{_{-0.5}}$\\
& & & & & & & & & \\
\hline
\end{tabular}
\end{table}
\end{document}